# FAR- AND MID-INFRARED SPECTROSCOPY OF COMPLEX ORGANIC MATTER OF ASTROCHEMICAL INTEREST: COAL, HEAVY PETROLEUM FRACTIONS, AND ASPHALTENES

**Franco Cataldo[1,2],  D. A. García-Hernández[3,4], Arturo Manchado[3,4,5]**

[1]INAF- Osservatorio Astrofisico di Catania, Via S. Sofia 78, Catania 95123, Italy
[2]Actinium Chemical Research srl, Via Casilina 1626A, Rome 00133, Italy
[3]Instituto de Astrofísica de Canarias, Vía Láctea s/n E-38200, La Laguna, Tenerife, Spain,
[4]Departamento de Astrofísica, Universidad de La Laguna (ULL), E-38205 La Laguna, Spain
[5]Consejo Superior de Investigaciones Científicas, Madrid, Spain

## Abstract

The coexistence of a large variety of molecular species (i.e., aromatic, cycloaliphatic and aliphatic) in several astrophysical environments suggests that unidentified infrared emission (UIE) occurs from small solid particles containing a mix of aromatic and aliphatic structures (e.g., coal, petroleum, etc.), renewing the astronomical interest on this type of materials. A series of heavy petroleum fractions namely "Distillate Aromatic Extract" (DAE), "Residual Aromatic Extract" (RAE), heavy aromatic fraction (BQ-1) and asphaltenes derived from BQ-1 were used together with anthracite coal and bitumen as model compounds in matching the band pattern of the emission features of proto-planetary nebulae (PPNe). All the model materials were examined in the mid-infrared (2.5-16.66 μm) and for the first time in the far-infrared (16.66-200 μm), and the infrared bands were compared with the UIE from PPNe. The best match of the PPNe band pattern is offered by the BQ-1 heavy aromatic oil fraction and by its asphaltenes fraction. Particularly interesting is the ability of BQ-1 to match the band pattern of the aromatic-aliphatic C-H stretching bands of certain PPNe, a result which is not achieved neither by the coal model nor by the other petroleum fractions considered here. This study shows that a new interesting molecular model of the emission features of PPNe are asphaltene molecules which are composed by an aromatic core containing 3-4 condensed aromatic rings surrounded by cycloaliphatic (naphtenic) and aliphatic alkyl chains. It is instead shown the weakness of the model involving a mixture of polycyclic aromatic hydrocarbons (PAHs) for modeling the aromatic infrared emission bands (AIBs). The laboratory spectra of these complex organic compounds represent a unique data set of high value for the astronomical community; e.g., they may be compared with the Herschel Space Observatory spectra (~51-220 μm) of several astrophysical environments such as (proto-) Planetary Nebulae, H II regions, reflection nebulae, star forming galaxies, and young stellar objects.

## 1. Introduction

The formation of elemental carbon in the Universe and its transformation into simple and complex molecules and macromolecules has been reviewed extensively (Tielens & Charnley 1997; Henning & Salama 1998; Kwok 2004, 2009, 2012; Cataldo 2004; Ehrenfreund & Cami 2010; Ehrenfreund & Foing 2010; Ehrenfreund et al. 2011). More than 160 different types of organic molecules are known in the interstellar and circumstellar medium including over 60 detected in the circumstellar medium of late-type carbon-rich stars (Kwok 2009, 2012; García-Hernández 2012). The largest organic molecules recognized to date in planetary nebulae are the fullerenes $C_{60}$ and $C_{70}$ (Cami et al. 2010; García-Hernández et al. 2010, 2011, 2012). The discovery of fullerenes was a success of the development of astronomical infrared spectroscopy (AIS) while smaller molecules were detected by radioastronomy. AIS has also detected a family of strong infrared emission bands at 3.3, 6.2, 7.7, 8.6, 11.3, and 12.7 μm (3030, 1613, 1299, 1163, 885 and 787 $cm^{-1}$) in a series of different astrophysical objects ranging from H II regions to reflection nebulae, and carbon-rich (proto-) planetary nebulae, and even in the diffuse interstellar medium. As reviewed by Kwok (2009) these bands are collectively known as aromatic infrared bands (AIBs). In most cases the AIBs are also accompanied by aliphatic infrared bands (e.g., those at 3.4, 6.9, and 7.3 μm) and other not yet assigned broad bands and plateaus, suggesting the presence of mixed aromatic-aliphatic organic nanoparticles (Kwok & Zhang 2011). These bands (both aromatic and aliphatic) are collectively known as unidentified infrared emission bands (UIE). In the nineties, also the far infrared spectral window (wavelengths longer than ~17 μm hereafter) was became accessible to AIS thanks to the Infrared Space Observatory (ISO) and another important emission feature has been observed at about 21 μm (476 $cm^{-1}$) in many carbon-rich proto-planetary nebulae (PPNe) (e.g., Volk et al. 1999) and assigned to a still unidentified solid-state carbonaceous carrier while it is not completely clear if another very broad feature at 30 μm (333 $cm^{-1}$) may be due to inorganic or organic compounds (see e.g., García-Hernández 2012 for a very recent review). The solid carbon dust from PPNe cannot be made by graphite as thought in the past but more reasonably by a complex and cross-linked molecular structure resembling coal. Papoular et al. (1996, 2001) as well as Guillois et al. (1996) were the first ones to bring our attention to coal and kerogen as a possible chemical model for understanding the chemical structure of cross-linked macromolecular networks found in PPNe and in other space environments. Indeed, certain types of mature or semi-mature coal are able to match the infrared emission of the aromatic infrared bands (AIBs) (e.g., Kwok 2012). Although the coal model could not be considered a perfect and definitive model, the alternative model of UIE is based on PAHs (e.g., Tielens 2008) and more precisely on PAH mixtures, which are able to match the AIBs pattern (e.g., Draine et al. 2007). Some years ago, Cataldo et al. (2002), Cataldo & Keheyan (2003), and Cataldo et al. (2004) have shown that also heavy petroleum fractions are able to match the band pattern of the UIE in PPNe and of coal in the mid infrared. In comparison to coal, the heavy petroleum fractions are viscous liquids at room temperature and become glassy solids below 235 K. The heavy petroleum fractions are not cross-linked as in the case of coal and have a mixed aromatic/aliphatic and naphthenic chemical structure. Since the PAHs model of UIE is not fully convincing there is a renewed interest in the coal model and also in the heavy petroleum fraction model of UIE (e.g., Kwok & Zhang 2011; García-Hernández 2012). Therefore, the present paper is dedicated to the exploration of the mid- and far-infrared (hereafter 2.5-16.66 μm and 16.66-200 μm for the mid- and far-infrared ranges, respectively) spectra of coal, a series of selected heavy petroleum fractions as well as asphaltenes derived from heavy petroleum fractions and from bitumen. The mid- and far-infrared spectra of certain acenes (e.g., tetrahydronaphthalene, anthracene, etc.) are explored as well.

## 2. Experimental

### 2.1 – Materials

A series of petroleum fractions which have given promising results in matching the band infrared pattern of the AIBs of certain PPNe in the mid-infrared (Cataldo et al. 2002, 2004; Cataldo & Keheyan 2003) were used also in the present work for the far-infrared spectroscopic analysis. The petroleum fractions used in the present work are known as DAE = distillate aromatic extract and T-RAE = treated aromatic extract, both supplied by ENI (Italy). The residue of atmospheric distillation of crude oil is distilled again under vacuum to produce distillate and residual lubricating oil base-stocks.

The distillate is extracted with polar solvents like furfural, dimethylsulphoxide or N-methylpyrrolidone to remove large part of the polar compounds present in the distillate. After the extraction, the solvent is distilled off and the refined distillate called DAE is obtained. The residue of vacuum distillation is first treated with liquid propane to remove resins and asphaltenes, and then is extracted with polar solvents to remove large part of the molecules containing heteroatoms. At the end of the process either a RAE or, if the refinement is push in two stages extraction, a T-RAE are obtained.

The oil fraction BQ-1 donated by Repsol (Spain) is a residue of vacuum distillation incorporating an important fraction of asphaltenes. A commercial bitumen sample supplied by ENI (Italy) was also used in the present work and the asphaltenes were recovered from the sample. A sample of standard anthracite coal (reference material no. 65) supplied by the Commission of the European Communities "Community Bureau of Reference BCR" with an individual identification analysis no. 417. Table 1 summarizes the physical and chemical properties of the oil fractions, asphaltenes, bitumen, and coal (see below).

**TABLE 1 - PHYSICAL AND CHEMICAL PROPERTIES OF THE OIL FRACTIONS, ASPHALTENES, BITUMEN AND COAL**

|  |  | DAE | RAE | T-RAE | BQ-1 | BQ-1 | BQ-1 | Bitumen | Bitumen | Coal |
|---|---|---|---|---|---|---|---|---|---|---|
|  |  |  |  |  |  | Deasphaltenized | Asphaltenes |  | Asphaltenes |  |
| Density | g/ml (15°C) | 0.990 | 0.944 | 0.932 | 0.995 | 0.985 |  | 1.020 |  | 1.698 |
| Viscosity | cSt 100°C | 26 | 51 | 25 | 88 | 63 | solid | solid | solid | solid |
| Softening point | °C | -38 | -45 | -40 | -26 | -32 |  | +40,8 |  | >365 d |
| Asphaltenes | % | 0 | 0 | 0 | 17 | 0 | 100 | 14 | 100 | n.d. |
| Aromatics | % | 40 | 27 | 25 | 55 | 42 | 80 | n.d. | 80 | n.d. |
| Naphthenics | % | 25 | 38 | 30 | n.d. | n.d. | n.d. | n.d. | n.d. | 6 |
| Paraffinics | % | 35 | 35 | 45 | n.d. | n.d. | n.d. | n.d. | n.d. | 4 |

Note: n.d. = not determined

A series of PAHs belonging to the acene class, namely naphthalene, anthracene, tetracene, and pentacene were purchased from Sigma-Aldrich as well as all the other PAHs mentioned in the present work, namely: phenanthrene, picene, pyrene, methylpyrene, perylene, acenaphte, acenaphtylene and 9,10-dihydroanthracene. The infrared spectra of these molecules were recorded in CsI matrix individually and as a mixture with a special attention to the feature at about 21 μm.

2.2 – Separation of petroleum asphaltenes from the oil fraction BQ-1

By definition the asphaltenes are the oil fraction insoluble in n-pentane, n-hexane, or n-heptane (Mullins et al. 2007). A BQ-1 sample of 5.90 g was mechanically stirred in a flask with 250 ml of n-hexane for 8 h at room temperature. The mixture appears turbid and was filtered through a paper filter. The asphaltenes remained as an insoluble light brown solid residue on the filter paper and were dried in a desiccator to constant weight. The asphaltene yield was 1.033 g; i.e., 17.5% of the starting weight. The deasphalted BQ-1 sample passed through the filter and was distilled under reduced pressure to strip off the solvent n-hexane leaving 4.85 g of dark brown oil.

2.3 – Separation of the bitumen asphaltenes from a bitumen sample

A bitumen sample of 2.139 g was dissolved in 80 ml of toluene and then mixed with 250 ml of n-hexane. The solution was mechanically stirred for 8 h at room temperature and then filtered through a paper filter. The insoluble asphaltenes from the bitumen sample were 0.302 g corresponding to 14.1% of the starting weight.

## 2.4 – Preparation of mixtures of PAHs for the far infrared spectroscopy

Each individual PAH of the acene series (naphthalene+anthracene+tetracene+pentacene) was weighted in amount of 1 mg and mixed in an agate mortar with the other acenes in the same weight proportion. Then CsI was added to the mixture, grinded and pressed in a pellet having a diameter of 8 mm and a thickness of 1.5 mm. The pellet was prepared using a Specac press at a pressure of 5 tonns. Similarly it was prepared the mixture of acenes and the asphaltenes of BQ-1 in the same weight proportions, the mixture of acenes and the asphaltenes both from BQ-1 and from bitumen in the same weight proportions, the mixture of all acenes with other PAHs (phenanthrene and picene, pyrene, methylpyrene, perylene, acenaphtene, acenaphtylene and 9,10-dihydroanthracene) in the same weight proportions.

## 2.5 – Methodology for FT-IR spectroscopy in the mid- and in the far-infrared

All the infrared spectra were recorded at room temperature on a Thermo-Scientific spectrometer Nicolet 6700. The mid-infrared spectra were recorded using a KBr beamsplitter and a DTGS/KBr detector. The spectra were recorded in transmittance mode and solid samples were embedded in KBr pellets while the liquid samples were closed as thin film between two KBr disks. The far-infrared spectra were obtained using a solid substrate beam-splitter and a DTGS/Polyethylene detector while the spectrometer was continuously flushed with dry air free from $CO_2$. The solid samples for far-IR spectroscopy were embedded in polyethylene powder while the liquid samples were included as thin film between two polyethylene windows. In some cases polyethylene was replaced by CsI as window material.

## 3. Results and Discussion

### 3.1 – Mid-infrared spectroscopy: asphaltenes and other petroleum fractions

In a series of previous works, Cataldo et al. (2002, 2004) and Cataldo & Keheyan (2003) have shown the effectiveness of certain petroleum fractions in matching the UIE band pattern of PPNe. The most promising oil fractions used in the previous works were distillate aromatic extract (DAE) and treated residual aromatic extracts (T-RAE) (Cataldo et al. 2004). They were used in pristine state or were chemically modified to change their aromatic content. Another interesting petroleum fraction was a heavy aromatic fraction called BQ-1 from Repsol (Cataldo & Keheyan 2003). These petroleum fractions are re-examined in the present work in the mid-infrared together with coal and bitumen samples and all samples are examined for the first time in the far-infrared. Another novelty of the current work is represented by the isolation and the analysis of petroleum and bitumen asphaltenes. Being sufficiently refined both DAE, RAE, and T-RAE oil fractions do not contain asphaltenes but the heavy aromatic oil fraction BQ-1 is composed by 17.5% by weight by asphaltenes and the bitumen sample is also constituted by 14.1% by asphaltenes.

Asphaltenes are interesting molecules from the astrochemical point of view because they are polycyclic aromatic hydrocarbons (PAHs) with alkyl substituents. More precisely, the alkyl substituents are aliphatic or cycloaliphatic in their nature. Mullins et al. (2007) have reported that asphaltenes from oil have an average molecular weight of 750 Da against the 500 Da of asphaltenes from coal. Furthermore, asphaltenes from petroleum are considerably more alkylated than those from coal which in their turn have a higher aromatic content. Asphaltenes from petroleum have only a single fused ring system per molecule and the alkyl chains have lengths comprised between 4 and 6 carbon atoms. The alkyl substitution of the polycyclic aromatic core of the molecule reduces its melting point so that the asphaltenes appear as low melting solids or viscous liquids at room temperature (Mullins et al. 2007). Another feature of petroleum asphaltenes regards the fact that each molecule may contain sulphur and nitrogen. On average, each petroleum asphaltene molecule has one atom of sulphur under the form of a thiophene structure or under the form of a sulphide/sulfoxide group. Instead, the nitrogen atom is generally under the form of a pyrrole or pyridine group. Another interesting property of asphaltenes regards the fact that the asphaltene molecules tend to stack each other under determined conditions because of the π-π interaction of their aromatic core.

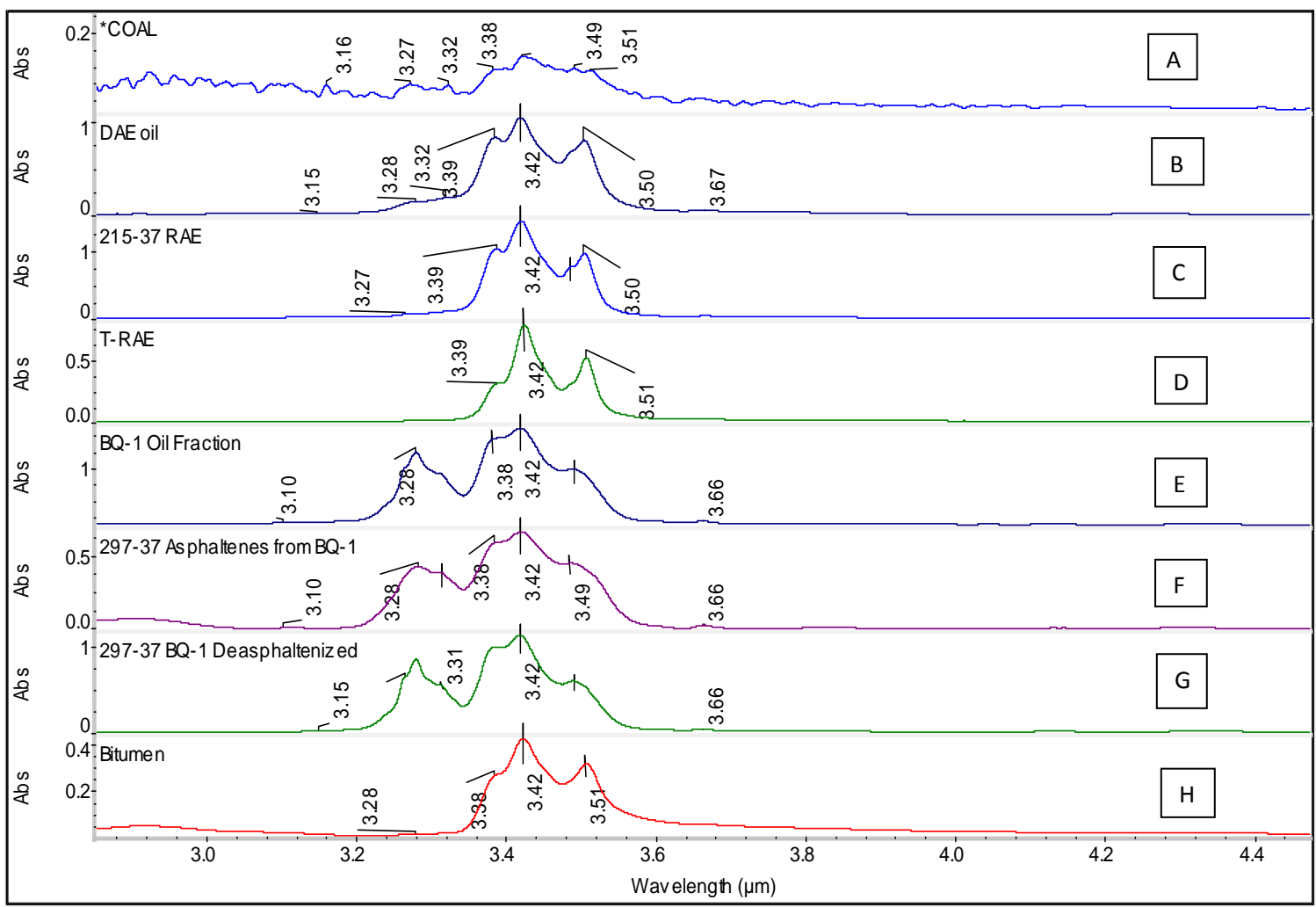

*Figure. 1 – FT-IR spectra in the C-H stretching band region. From top to bottom are shown respectively the spectra of **A** = Coal; **B** = DAE; **C**= RAE; **D** = T-RAE; **E** = BQ-1; **F** = Asphaltenes from BQ-1; **G** = Deasphaltenized BQ-1; **H** = Bitumen.*

The C-H stretching bands of all materials examined in the present work are displayed in Figure 1. The typical aromatic C-H stretching band is expected at about 3.3 μm. Indeed, the most intense infrared band at 3.28 with an important shoulder at 3.31 μm can be observed both on the heavy petroleum fraction BQ-1, on its asphaltenes and even on the deasphalted BQ-1 sample, implying that the removal of solid, n-hexane insoluble asphaltenes from BQ-1 has reduced slightly its aromaticity and hence also the smaller molecules of deasphalted BQ-1 have a similar chemical structure of the larger removed as asphaltenes. On the other hand, the other petroleum fractions DAE, RAE, and T-RAE show only weak to negligible absorption bands in the aromatic C-H stretching region, respectively, but this is compensated by the aliphatic bands centered at 3.39, 3.42 and 3.50 μm, suggesting that the polycyclic aromatic "core" of the molecule is heavily alkylated. BQ-1 and its asphaltenes combine an intense aromatic C-H stretching band with important infrared bands in the aliphatic and naphthenic C-H stretching region suggesting a less alkylated aromatic "core" than the DAE, RAE and T-RAE fractions. The infrared bands at 3.42 and 3.50 μm are due respectively to asymmetric and symmetric $CH_2$ stretching while the presence of $CH_3$ groups is suggested by the 3.38 μm band. A general feature among the different organic compounds analyzed is that the aromatic C-H stretching band at 3.3 μm is always less intense than the aliphatic one at 3.42 μm, possibly as a consequence of the high aliphatic content in our sample of model compounds. In addition, it is to be noted here that the CH stretching infrared bands are the most intense ones in several organic compounds in our sample such as bitumen, DAE, RAE, and asphaltenes from BQ-1 (see Section 4). In any case, for comparison, in Figure 2 (updated from Geballe et al. 1992 and courtesy of S. Kwok) we display the observed emission bands in the C-H stretching spectral region of the three PPNe IRAS 04296+3429, IRAS 22272+5435, and CRL 2688 (see also Hrivnak et al. 2007). IRAS 04296+3429 and IRAS 22272+5435 display the aliphatic 3.42 μm feature as intense as the aromatic 3.3 μm feature. However, the post-AGB star CRL 2688 shows some of these emission bands with a neat predominance of the emission at 3.3 μm (Figure 2). Thus, in this spectral region the best laboratory resemblance is offered by BQ-1 and its fractions (where the 3.3 and 3.4 μm features are of similar intensity) rather than the other oil fractions. Even the spectrum of anthracite coal in this region is somewhat faint with a predominance of the aliphatic absorption bands; the very weak aromatic C-H stretching is distinguishable at 3.27 and 3.32 μm.

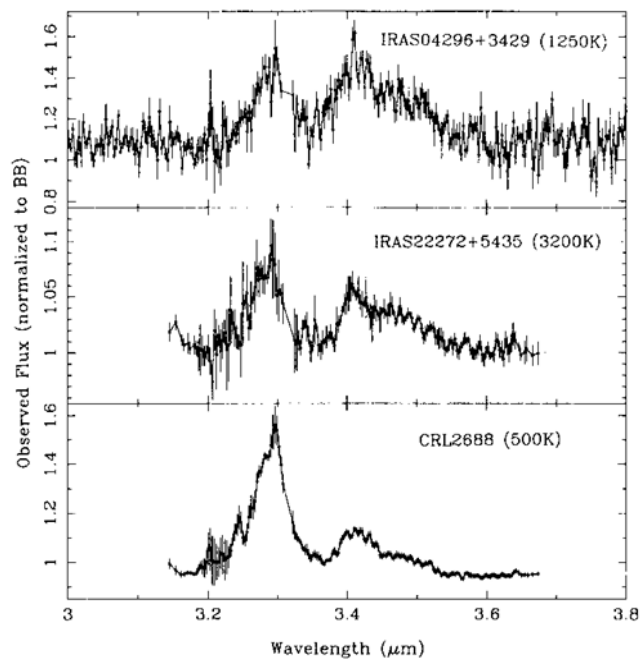

*Figure 2. Infrared spectra in the CH stretching region (~3-4 μm) of the PPNe (from top to bottom) IRAS 04296+3429, IRAS 22272+5435, and CRL 2688 (updated from Geballe et al. 1992 and courtesy of S. Kwok, 2011; see also Hrivnak et al. 2007).*

In the spectral region between 4.5 and 8.0 μm the main UIE in young PNe and PPNe are located at 6.2, 6.9, and 7.7 μm (Hrivnak et al. 2000; Kwok 2012). Peeters et al. (2002) have used the position of the AIBs at 6.2, 7.7, and 8.6 μm to classify the astrophysical objects with these bands into three classes. In Figure 3, it is possible to observe that coal gives a very broad band at 6.24 µm while the petroleum fractions DAE, RAE, and T-RAE show a weak band. Instead, the fraction BQ-1 and its derivatives give a neat and clear band at 6.26 µm accompanied by another peak at 6.63 µm. Figure 3 shows also, for all materials examined, a broad feature at 6.88 µm with a shoulder at 6.95 µm, hence coincident with the discrete aliphatic band at 6.90 µm. The other important band in terms of intensity in the spectra of Figure 3 occurs at 7.27 µm for all samples analyzed and only for BQ-1 and derivatives occurs as a doublet at 7.20 and 7.26 µm. The band a 7.65 µm in the spectra of Figure 3 is rather weak in all cases but the oil fraction BQ-1 displays also a band at 7.89 µm. Both these bands are in relation with the AIB at 7.7 µm but our laboratory spectra cannot explain the AIB at this wavelength. Surprisingly, in Figure 4 the oil BQ-1 and its derivatives show a weak band at about 8.6 µm while such band is much more intense and broad in coal; being also weak in the other oils DAE, RAE, and T-RAE. It is to be noted here that a very similar pattern of UIE at 6.2, 6.6, 6.9, and 7.3, and 7.7-7.9 µm together with the lack of the AIB at 8.6 µm has been very recently observed in a few very young PNe with fullerenes (García-Hernández et al. 2011, 2012). The set of three UIE features composed by the 6.6, 6.9, and 7.3 µm is very unusual and the BQ-1 oil is the unique model compound showing these three features. In principle, it is difficult that the BQ-1 oil fraction can completely explain these three features in fullerene-containing young PNe – mainly because the stronger 12.7 µm complex seems to be absent in the astronomical sources – but some contribution may explain the variable strength of these features from source to source.

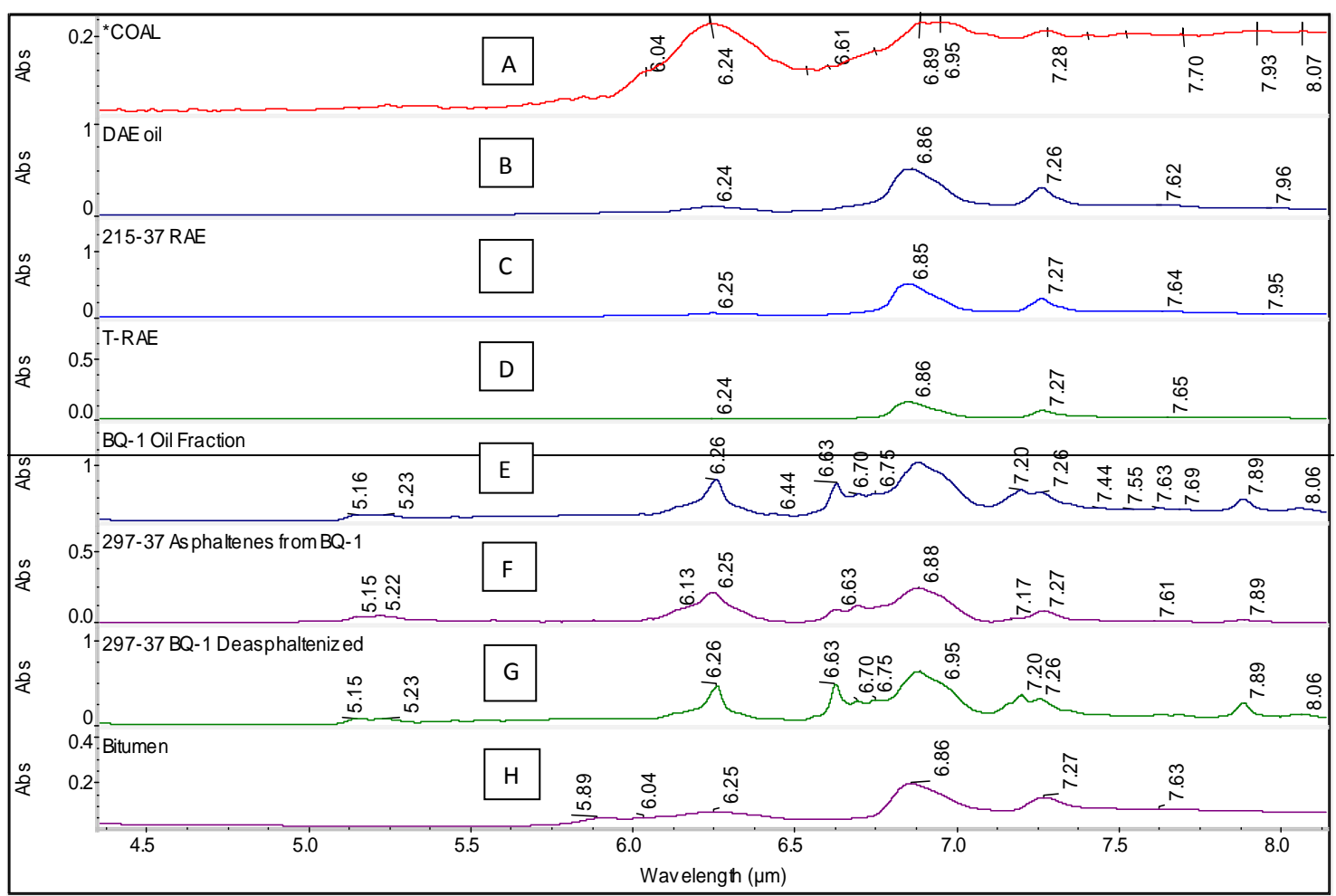

*Figure 3 – FT-IR spectra between 4.5 and 8.0 µm. From top to bottom are shown respectively the spectra of* **A** = Coal; **B** = DAE; **C**= RAE; **D** = T-RAE; **E** = BQ-1; **F** = Asphaltenes from BQ-1; **G** = Deasphaltenized BQ-1; **H** = *Bitumen.*

**TABLE 2 - COMPARISON OF MID AND FAR-IR SPECTRA OF MODEL COMPOUNDS WITH PPNe AND OTHER SOURCES(ALL DATA IN um)**

| COAL | DAE | RAE | T-RAE | BQ-1 | ASPHALTENES FROM BQ-1 | DE-ASPHALTENIZED | BITUMEN | Kwok 2012 p.139 PPNe gen | Kwok 2012 p 103 IRAS 21282+5050 Young PNe | Kwok 2012 p.104 IRAS 22272+5435 PPNe | Kwok 2012 p.128 NGC7023 refl nebula | Smith et al. 2007 Tab 3 Star Forming galaxies |
|---|---|---|---|---|---|---|---|---|---|---|---|---|
| 3.16 | 3.15 | 3.27 | | 3.10 | 3.10 | 3.15 | | | | | | |
| 3.27 | 3.28 | | | 3.28 | 3.28 | 3.28 | 3.28 | | | | | |
| 3.32 | 3.32 | | | 3.31 | 3.31 | 3.31 | | 3.3 | 3.29 | | 3.3 | |
| 3.38 | 3.39 | 3.39 | 3.39 | 3.38 | 3.38 | 3.38 | 3.38 | | | | | |
| 3.42 | 3.42 | 3.42 | 3.42 | 3.42 | 3.42 | 3.42 | 3.42 | 3.4 | 3.4 | 3.4 | | |
| 3.49 | | | | | 3.49 | 3.49 | | | 3.46 | | | |
| 3.51 | 3.50 | 3.50 | 3.51 | 3.51 | | | 3.51 | | 3.51 | | | |
| | 3.67 | 3.67 | | 3.67 | | | | | 3.56 | | | |
| | | | | 5.16 | 5.16 | 5.16 | | | | | | |
| | | | | 5.23 | 5.23 | 5.23 | | | | | | 5.27 |
| | | | | | | | 5.89 | | | | | 5.7 |
| | | | | | | | 6.04 | | | | | |
| 6.24 | 6.24 | 6.25 | 6.24 | 6.26 | 6.26 | 6.26 | 6.25 | 6.2 | 6.2 | 6.2 | 6.2 | 6.22 |
| 6.61 | 6.68 | | | 6.63 | 6.63 | 6.63 | | | | | | |
| | | | | 6.7 | | 6.7 | | | | | | 6.69 |
| | | | | 6.75 | | 6.75 | | | | | | |
| 6.89 | | 6.85 | 6.86 | 6.86 | 6.88 | 6.86 | 6.86 | 6.9 | | 6.9 | 6.9 | |
| 6.95 | | | | | | 6.95 | | | | | | |
| 7.28 | 7.26 | 7.27 | 7.27 | 7.26 | 7.27 | 7.26 | 7.27 | | | | | |
| | | | | 7.44 | | | | | | | | 7.42 |
| 7.52 | | | | 7.55 | | | | | | | | |
| | 7.62 | 7.64 | 7.65 | 7.63 | 7.61 | | 7.63 | 7.7 | 7.7 | 7.7 | 7.5 | 7.60 |
| 7.7 | | | | 7.69 | | | | | | | | |
| | | | | 7.89 | 7.89 | 7.89 | | | | | 7.8 | |
| 7.93 | 7.96 | 7.95 | | | | | | | | | | |
| 8.07 | | | | 8.06 | 8.06 | 8.06 | | | | | | |
| | | | | 8.28 | 8.28 | 8.26 | | | | | | 8.33 |
| 8.62 | 8.60 | 8.60 | 8.67 | 8.6 | 8.66 | 8.6 | 8.59 | 8.6 | 8.6 | | 8.6 | 8.61 |
| 8.94 | | | | 8.88 | 8.88 | 8.88 | | | | | | |
| 9.28 | | | | 9.32 | 9.32 | | | | | | | |
| | 9.45 | | | | 9.70 | 9.69 | 9.69 | | | | | |
| 9.68 | | 9.68 | 9.68 | 9.9 | 9.81 | 9.9 | | | | | | |
| | 10.43 | 10.39 | 10.38 | 10.42 | 10.40 | 10.44 | 10.40 | | | | | |
| 10.64 | | | | 10.58 | 10.58 | | | | | | | 10.68 |
| | | | | 10.98 | | 10.95 | | | | | | 11.23 |
| 11.57 | 11.54 | 11.57 | 11.54 | 11.39 | 11.42 | 11.39 | 11.54 | 11.3 | 11.3 | 11.4 | 11.3 | 11.33 |
| | | | | 11.83 | 11.88 | 11.81 | | | | | | 11.99 |
| 12.37 | 12.34 | 12.33 | 12.31 | 12.29 | 12.27 | 12.31 | 12.35 | 12.4 | | 12.1 | 12.1 | |
| | | | | 12.77 | 12.78 | 12.78 | | 12.7 | 12.4 br | | 12.7 | 12.62-12.69 |
| | | | | | | 12.97 | | | | | | |
| | 13.10 | 13.13 | 13.11 | | | | | | | | | |
| 13.41 | 13.34 | 13.38 | 13.49 | 13.51 | 13.32 | 13.53 | 13.37 | 13.3 | 13.40 | | | 13.48 |
| | 13.73 | 13.82 | 13.87 | | | 13.93 | 13.84 | | | | | 14.04 |
| | 14.27 | 14.30 | | 14.29 | 14.29 | 14.29 | | | | 14.2 | | 14.19 |
| 14.90 | 14.97 | | | | | 14.96 | 14.86 | | | | | |
| 15.37 | | | | | | | 15.39 | | | | | |
| 16.01 | | | | 16.16 | 16.16 | 16.15 | | | | | | 15.90 |
| 16.58 | 16.25 | 16.74 | | | | 16.88 | | | 16.8 br | | 16.40 | 16.45 |
| 17.02 | | | | | | | | 17.0 | | | | 17.04 |
| 17.47 | 17.39 | 17.29 | | | | 17.34 AS* | | 17.4 | | | 17.4 | 17.37 |
| 17.86 | 17.81 | 17.77 | | | | | 17.85 | 17.8 | | | | 17.87 |
| 18.44 | | | | 18.2 | 18.4 | | | | | | | |
| 18.59 | 18.52 | | | | 18.77 | 18.88 | | | | | 18.9 | 18.92 |
| | | | | 18.75 | 19.04 | | | | | | | |
| | | | | | | 20.28 | | | | | | |
| 21.30 | 21.00 | 21.00 | | 21.02 | 21.02 | 21.06 | | 21.00 | 21.00 | 20.3 | | |
| | | | | | | | 21.60 AS* | | | | | |
| 23.50 | 23.32 | 23.25 | | 23.99 | | 23.49 | | | | | | |
| | | | | | | 24.54 | | | | | | |
| | | | | 27.82 | | | | | | 26.0 | | |
| 29.76 | 29.22 | | | | | | | 30.0 | | | | |
| 33.31 | | | | | | | | | | | | 33.10 |
| 37.38 | 38.89 | 39.40 | | 39.71 | 41.54 | | | | | | | |
| 42.18 | | | | | | | | | | | | |
| 51.99 | | | | | | | | | | | | |
| 56.62 | | 56.37 | | 55.69 | | | | | | | | |
| 61.06 | | | | | | | | | | | | |
| | | 66.66 | | | | | | | | | | |
| 73.27 | | | | | | | | | | | | |
| 80.37 | | | | | | | | | | | | |
| 90.80 | | | | | | | | | | | | |
| 96.23 | | | | | | | | | | | | |
| 110.33 | | | | | | | | | | | | |
| 117.83 | | | | | | | | | | | | |
| 128.58 | | | | | | | | | | | | |
| 142.98 | | | | | | | | | | | | |
| 152.90 | | | | | | | | | | | | |
| 167.30 | 162.10 | | | | | | | | | | | |
| 185.35 | 185.14 | | | | | | | | | | | |
| | | | | (AS* = MEASURED ON BITUMEN ASPHALTENES) | | | | | | | | |

Another relatively intense band is shown in Figure 4 at 9.68 µm for all materials examined and due to the C-H in-plane bending. Only BQ-1 and derivatives shows a couple of very weak bands in this region at 9.70 and at 9.90 µm. Then in Figure 4 we can observe a series of out-of plane aromatic C-H bending bands. The isolated C-H bending occurs at 11.57 µm for all materials with the exception of BQ-1 and derivatives which show two bands at 11.42 and 11.80 µm which may implies two different chemical structures having an isolated aromatic C-H bending. Typical AIBs in PPNe occurs at 11.3 µm while in star forming galaxies there are several bands like those at 11.23, 11.33, and 11.99 µm (Smith et al. 2007). The out-of-plane bending of two adjacent aromatic CH groups occurs at 12.37 µm (Colthup et al. 1990) and this time all samples in Figure 4 display such a feature in line also with the AIB of PPNe at 12.3-12.4 µm (Kwok 2012 p. 139). Only BQ-1 and derivatives again display clearly also the additional out of plane bending of three adjacent hydrogen atoms at 12.78 µm while the AIB in PPNe is normally located at 12.3 µm (Kwok 2012 p. 139) although some PPNe show a feature just at 12.7 µm (see Table 2). Four adjacent hydrogen bending is suggested by the band at 13.41 µm for all samples in Figure 4 with the exclusion of BQ-1 and its deasphalted derivative, which show this band located at 13.51 µm. Instead, the asphaltenes from BQ-1 show this band at 13.32 µm. The young PN IRAS 21282+5050 (Kwok 2012 p.103, see Section 4) shows this band at 13.4 µm. Five adjacent CH out of plane bending are observed between 13.70 and 14.20 µm (Colthup et al. 1990) and indeed such feature is shown by all samples with the exception of coal. A feature at 14.2 µm is observed also in the young PN IRAS 21282+5050 (Kwok 2012 p.103). In Figure 4 the band at 14.29 µm is more pronounced just in the BQ-1 sample and derivatives and this is logically due also to the high intensity of the C-H stretching band which is unique among the samples examined. Finally, Figure 4 shows also a feature at 16.16 µm for BQ-1 and derivatives, which can be connected with the band at 15.90 µm recorded in star forming regions (see Table 2) and with the unidentified 15.8 µm feature observed in certain PPNe (Kwok 2012 p.139).

All the mid-infrared bands of the petroleum fractions, coal, and bitumen are summarized in Table 2 together with the infrared emission bands seen in PPNe, reflection nebulae, and star forming galaxies.

3.2 – Far-infrared spectroscopy: asphaltenes and other petroleum fractions

Far-infrared spectroscopy has recently been applied to a series of common (Boersma et al. 2011) and less common PAHs (Mattioda et al. 2009) in relation to the AIBs and in particular to a couple of features observed in the far-infrared emission spectra of PPNe at about 21 and 30 µm (see Section 1). In Figure 5 we have displayed all the petroleum fractions, the oil and bitumen asphaltenes, and coal in the far infrared up to 80 µm. It is to be noted here that all laboratory spectra are almost featureless from 80 to 200 µm. Only coal and DAE spectra show some bands in the spectral region from 80 to 200 µm (Table 2) but these far-IR bands seem to be much weaker than those in the mid-IR. All laboratory data are also reported in Table 2. The far-infrared region of aromatic compounds is dominated by the out-of-plane quadrant ring bend vibrations (Colthup et al. 1990) and, more in detail, Mattioda et al. (2009) and Boersma et al. (2011) distinguish the "Jumping Jack" modes between 20 and 130 µm (where the molecules deform in-plane around its central core and symmetrically around one axis) from the "Drumhead mode" beyond 130 µm (where the PAH "sheet" makes out-of-plane wave motion similar to the membrane of a percussion instrument) and the "Buttefly" modes (where the molecules makes a synchronized out-of-plane flapping motion around a central axis of symmetry).

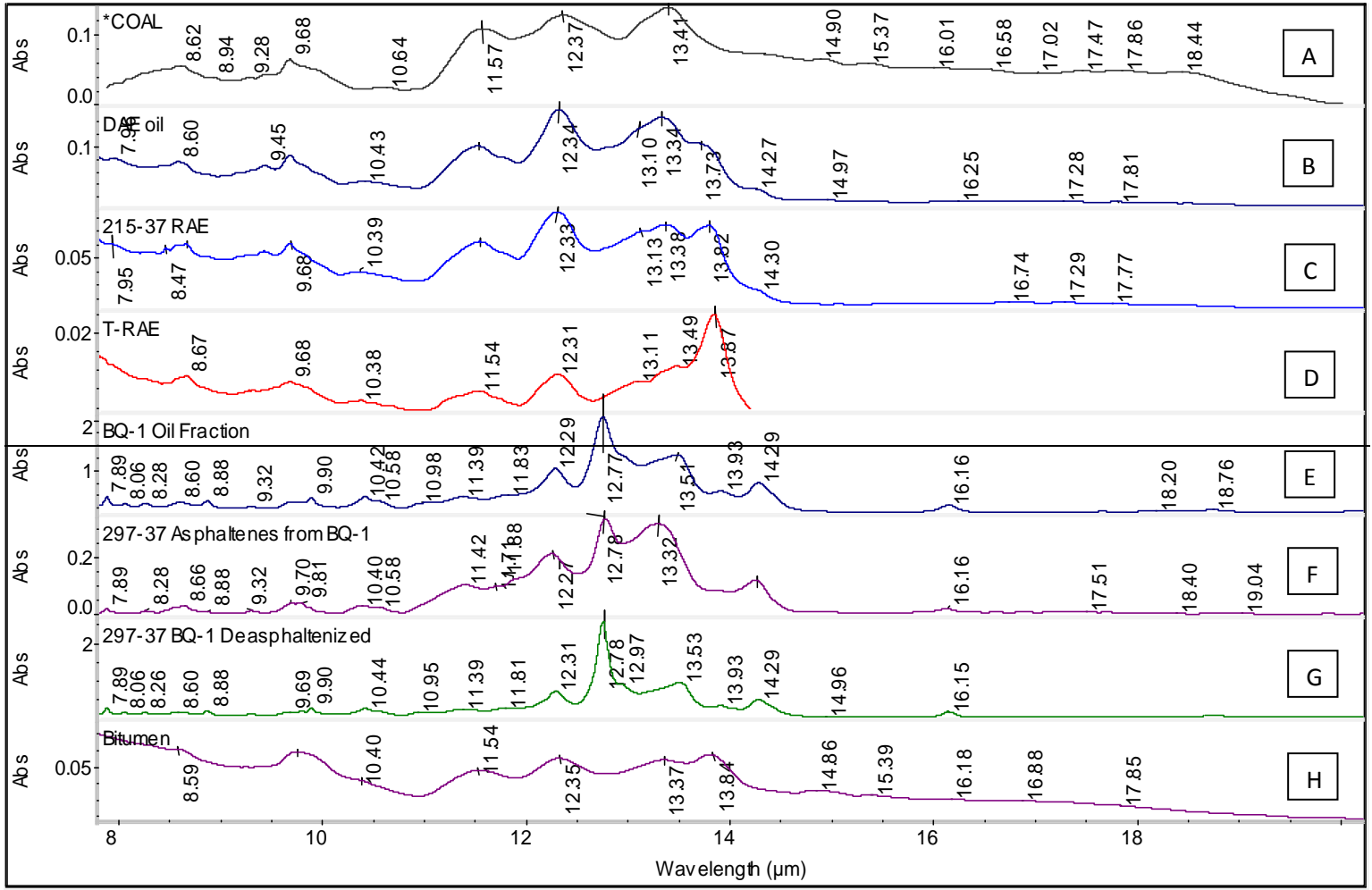

*Figure 4 – FT-IR spectra between 8.0 and 20.0 µm. From top to bottom are shown respectively the spectra of* ***A*** *= Coal;* ***B*** *= DAE;* ***C****= RAE;* ***D*** *= T-RAE;* ***E*** *= BQ-1;* ***F*** *= Asphaltenes from BQ-1;* ***G*** *= Deasphaltenized BQ-1;* ***H*** *= Bitumen.*

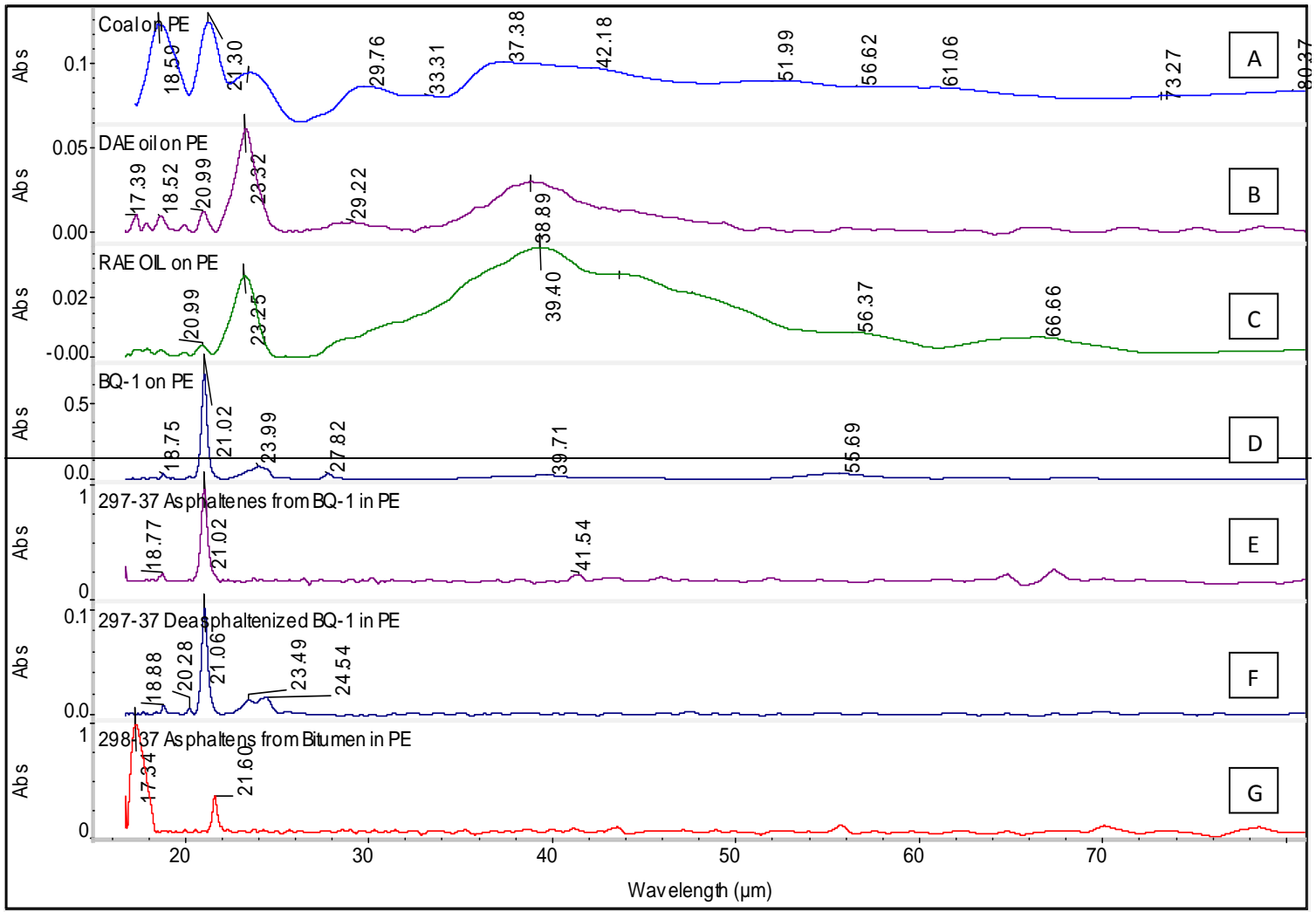

*Figure 5 − FT-IR spectra in the FAR-INFRARED between 16.0 and 80.0 µm. From top to bottom are shown respectively the spectra of **A** = Coal; **B** = DAE; **C**= RAE; **D** = BQ-1; **E** = Asphaltenes from BQ-1; **F** = Deasphaltenized BQ-1; **G** = Asphaltenes from bitumen.*

As expected the far-infrared spectrum of coal appears very rich of bands. Figure 5 shows the most important at 18.50, 21.30, 23.32, 29.76, 37.38, and 51.99 µm, followed by a series of other bands which are covering the entire far-infrared spectral range as summarized in Table 2. The richness of bands observed in the far infrared spectrum of coal, however, has no counterpart in astronomical observations and this can be considered a weakness of the coal model. Furthermore, such complex coal spectrum in the far-infrared was expected as a consequence of the contribution of different types of PAH substructures and substituents (Boersma et al. 2011). The petroleum fractions DAE and RAE, which in the mid-infrared have a spectrum somewhat similar to that of coal, in the far-infrared show the main band at 23.32 and 23.35 µm, respectively (see Figure 5), preceded by a series of weak bands (see Table 2) and followed by a broad band respectively at 38.89 and at 39.40 µm (Figure 5). The latest feature corresponds to the bands at 37.38 and 42.18 µm of coal. However, the far-infrared spectra of DAE and RAE are much less rich in far-infrared bands than coal, suggesting a more uniform and simpler average molecular structure. An interesting result is offered by the BQ-1 fraction and the far-infrared spectra of its asphaltene and deasphalted fractions (Figure 5 and Table 2). In all cases, there is a main band that is centered at 21.0 µm, relatively near of the feature observed in the emission spectra of PPNe (e.g., Volk et al. 1999). However, the astronomical feature seems to be almost invariant at a position of 20.3 µm (e.g., Hrivnak et al. 2009), being stronger and broader than the feature seen in the lab. The asphaltenes from BQ-1 show just two other weak features at 18.77 and 41.54 µm, while the pristine BQ-1 fraction shows also a broad band at 23.99 µm which appears split in two features in the deasphalted BQ-1, i.e., at 23.49 and 24.54 µm. Another interesting case of study is offered by bitumen: the pristine bitumen sample did not show any far-infrared emission band but the asphaltenes isolated from bitumen show an interesting spectrum reported in Figure 5 and characterized by two bands at about 17.34 µm that are not found in any other of the analyzed petroleum fractions and a feature at 21.60 µm, somewhat shifted at a longer wavelength than the 21.0 µm feature of BQ-1 and derivatives but not at about 23.3 µm as observed in DAE and RAE oils.

The final judgment on the validity or utility of the coal model against the petroleum fraction models will be played in the far infrared spectral region when more and more astronomical observations in this spectral region will become available in the future (e.g., by the Herschel Space Observatory). In this paper we make available the far-IR spectra of these complex organic compounds for the first time.

4 – Infrared spectra of certain PAHs in comparison with the petroleum fractions

A complete survey of the far-infrared spectra of PAHs was published by Boersma et al. (2011) while for coronene, dicoronylene, and ovalene one can take into consideration the work of Mattioda et al. (2009). From the computer-calculated spectra of Boersma et al. (2011) it appears evident that only the acenes, i.e., the PAHs having linearly condensed benzene rings, display exclusively the far-IR band at 21.0 µm without any additional important feature in the far-IR up to >200 µm. It is important to underline that only the acenes in the neutral state have a rather simple far-IR spectrum with only the feature at about 21.0 µm because the far-IR spectrum of their cations or their anions is already more complex, with additional strong bands in the far-IR or with a shift of the main band originally at 21.0 µm for neutral acenes to longer wavelengths for the corresponding cations. Figure 6 shows our experimental spectra of a series of neutral acenes (see Figure 7) and polystyrene. The latter shows an intense and broad peak at 18.52 µm and is reported just for reference being not an acene but a linear polymer with pendant phenyl groups. Apart the case of naphthalene that in addition to the band at 20.75 µm (accompanied by a small band at 21.18 µm) display also other bands in the far-IR at 27.59 and 56.99 µm, all the other acenes show a unique main band in the far-IR at about 21.0 µm. In fact, anthracene shows its main band at 21.08 µm (accompanied by a secondary band at 21.52 µm), tetracene has its main band at 21.27 µm (accompanied by a secondary band at 21.89 µm) and pentacene displays its main band at 21.44 µm (accompanied by a secondary band at 22.08 µm). There is slight regular shift of the main band of acenes towards longer wavelengths as a consequence of the addition of another ring. However, such shift becomes smaller and smaller as the acene becomes bigger. Indeed, from naphthalene to anthracene the shift is 0.33 µm, from anthracene to tetracene 0.19 µm, and from tetracene to pentacene 0.17 µm; so that it reasonable to predict that even bigger acenes than pentacene may display a similar spectrum of the latter. Since the petroleum fraction BQ-1 and its derivatives, both asphaltenes and the deasphalted fraction display a unique far-IR band at 21.0 µm without any additional feature up to 200 µm (see Table 2), it can be anticipated that we can take the acenes as model compound for the core of the molecules composing the BQ-1 fraction, which indeed have also an aliphatic and cycloaliphatic fraction in their chemical structure.

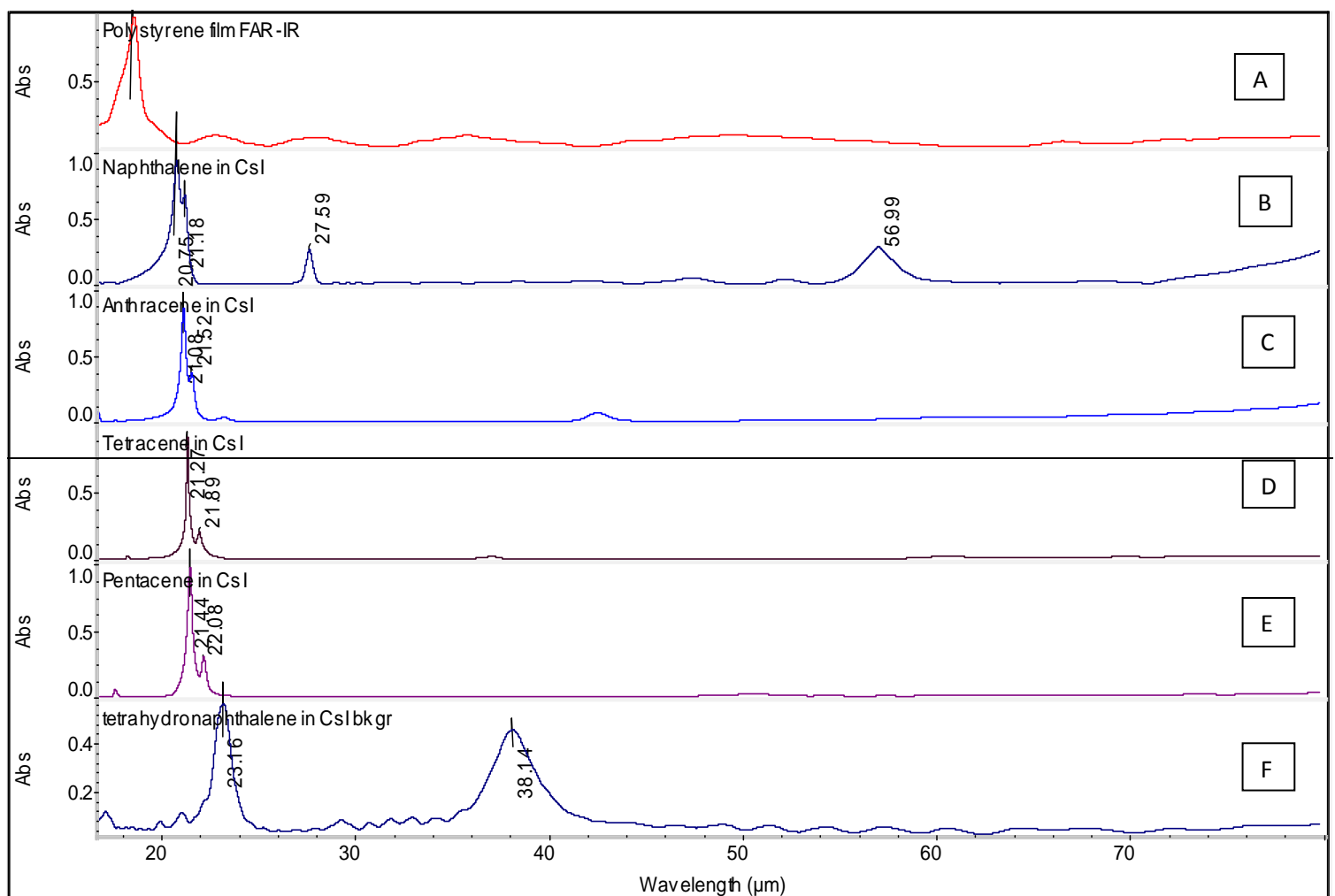

*Figure 6 – FT-IR spectra in the FAR-INFRARED between 16.0 and 80.0 μm of a series of acenes with the exclusion of the first spectrum on top which is due to polystyrene (**A**). The latter is followed from top to bottom by the spectrum of **B**= naphthalene, **C**= anthracene, **D**= tetracene, **E**= pentacene and **F**= tetrahydronaphthalene.*

It is for this reason that in Figure 6 (at the bottom) is reported the far-IR spectrum of tetrahydronaphthalene. This is an interesting molecule with two condensed hexagonal carbon rings (see Figure 7). One ring is fully aromatic and the other ring is completely hydrogenated. Therefore, tetrahydronaphthalene is 50% aromatic and 50% cycloaliphatic (or naphthenic as said by chemists). The far-IR spectrum of tetrahydronaphthalene is composed by two bands one located at 23.16 μm (shifted 2.41 μm towards longer wavelengths in comparison to naphthalene) followed by a new broad band at 38.14 μm. This implies that the partial hydrogenation of acenes causes the growth of a new band in the far-IR and a shift of the main band originally located at about 21 μm.

It is very interesting to compare the far-IR spectrum of tetrahydronaphthalene (Figure 6, at the bottom) with the far-IR spectra of the heavy petroleum fractions DAE and RAE shown in Figure 5 (see also the comparison in Table 3). The far-IR spectra of the DAE and RAE petroleum fractions display two main bands: a narrow band at 23.32 and 23.25 μm, respectively, and another broad one at 38.89 and 39.40 μm, respectively. These bands should be compared with the bands of tetrahydronaphthalene, a narrow one at 23.16 μm and a broader band at 38.14 μm; the spectral analogy is impressive. Table 3 shows also that DAE and RAE have also analogies with 9,10-dihydroanthracene. This may suggests that the aromatic "core" of these molecules may be based on the tetrahydronaphthalene chemical structure with some minor contribution from the structure of 9,10-dihydroanthracene. In Figure 8, we display the molecular models of the DAE and RAE petroleum fractions. Being molecular models, they wish to illustrate and suggest the possible structure of the "average" molecule in the petroleum fraction. Of course, DAE and RAE in reality are extremely complex mixtures of a number of different molecules. For a discussion of the molecular models in petroleum chemistry and about the "average" molecule we refer the reader to Cataldo et al. (2002, 2003, 2004).

| TABLE 3 - COMPARISON OF FAR-IR SPECTRA OF MODEL PAHs AND H-PAHs WITH PETROLEUM FRACTIONS (ALL DATA IN µm) | | | | | | | | | | | |
|---|---|---|---|---|---|---|---|---|---|---|---|
| Naphthalene | | | **20.75** | 21.18 | | 27.59 | | | | | 56.99 |
| Anthracene | | | **21.08** | 21.58 | | | | | 42.58 | | |
| Tetracene | 18.2 | | **21.27** | 21.89 | | | | 37.7 | | | |
| Pentacene | 18.0 | | **21.44** | 22.08 | | | | | | | |
| **BQ-1** | 18.75 | | **21.02** | | 23.99 | 27.82 | | 39.71 | | | 55.69 |
| **BQ-1 asphaltenes** | 18.77 | | **21.02** | | | | | | 41.54 | | |
| **BQ-1 deasphalted** | 18.88 | 20.28 | **21.06** | | 23.49 | 24.54 | | | | | |
| | | | | | | | | | | | |
| Tetrahydronaphthalene | | | | | 23.16 | | | 38.14 | | | |
| 9,10-Dihydroanthracene | | 20.03 | (21.07*) | (21.55*) | 22.90 | 25.86 | 33,69 | 37.39 | (42.54*) | 44,41 | 54.03 |
| **DAE** | | 20.99 | | | 23.32 | | | 38.89 | | | |
| **RAE** | | 20.99 | | | 23.25 | | | 39.40 | | | 56.37 |

(*) Anthracene impurity in 9,10-dihydroanthracene

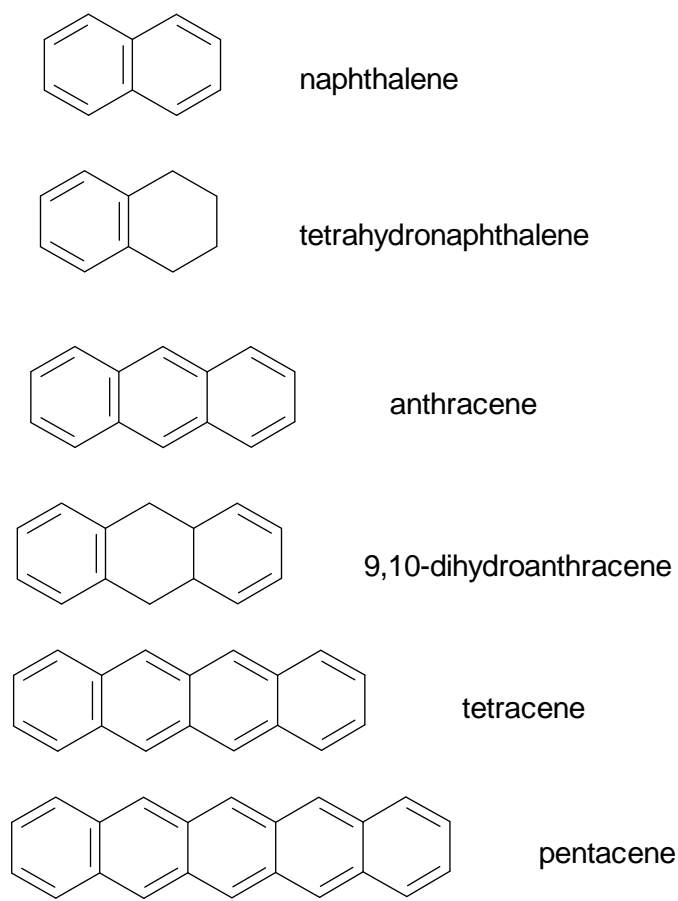

naphthalene

tetrahydronaphthalene

anthracene

9,10-dihydroanthracene

tetracene

pentacene

*Figure 7 – Chemical structure of several acenes.*

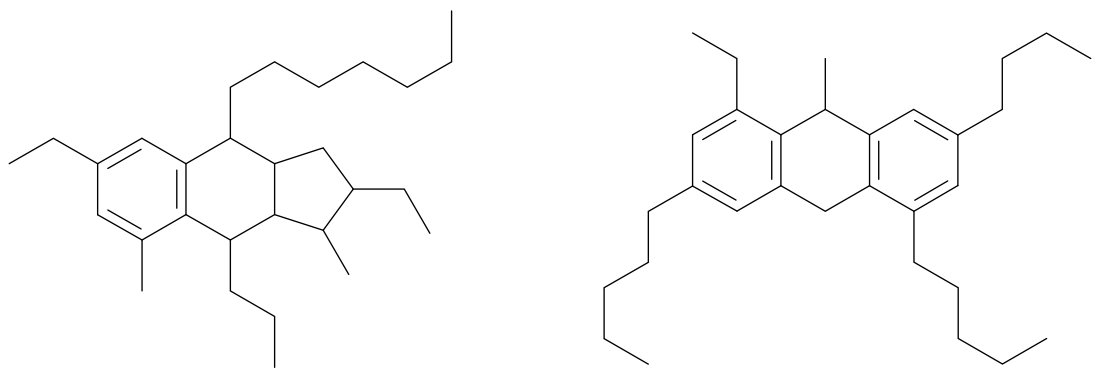

*Figure 8 – Petroleum molecular models of DAE and RAE petroleum fractions with tetrahydronaphthalene and 9,10- dihydroanthracene "core". The basic chemical structures were taken from Cataldo et al. (2004) and slightly modified to take into account the current experimental spectral data collected in the far infrared.*

As we have mentioned before, the heavy aromatic BQ-1 petroleum fraction and its derivatives asphaltene and deasphalted fractions display a far-IR spectrum (see Figure 5) which is surprisingly similar to the far-IR spectrum of the acenes (Figure 6; see also Table 3). Especially for the asphaltene fraction of BQ-1 (which is the most aromatic fraction), the unique main feature is at 21 µm with another small band at 41.54 µm, very close to the anthracene feature at 42.58 µm. The pristine BQ-1 and the deasphalted BQ-1 instead show additional features which are typical of the hydrogenated acenes, for instance at about 24 µm and at about 42 and 55 µm (see Table 3). Thus, the "core" of the molecules of BQ-1 asphaltenes is of anthracene type with some minor contribution from the tetracene structure. Some examples of molecular models or "average" molecules of the BQ-1 asphaltenes are shown in Figure 9. Instead the pristine BQ-1 and deasphalted BQ-1 - being less aromatic than the asphaltene fraction - show also spectral features attributable to 9,10-dihydroanthracene and tetrahydronaphthalene. In the recent years, a great progress has occurred in our understanding of the chemical structure of asphaltenes (Mullins et al. 2007). It has been found that indeed the petroleum asphaltene molecules are constituted by 3-4 condensed aromatic rings with a series of cycloaliphatic and aliphatic moieties in line with our results derived from FT-IR spectroscopy in the mid- and far-infrared. It must be emphasized that thanks to the application of the far-infrared spectroscopy of asphaltenes, we have determined the "core" of the asphaltene molecules. The asphaltene molecules are not only alkylated, they contain also some heteroatoms as shown in the Figure 9 (Mullins et al. 2007). In fact, knowing the "core" of these molecules as suggested by far-infrared spectroscopy and combining these results with our knowledge of asphaltene molecules in Figure 9 we have drawn some reasonable model molecule of the petroleum fractions, which represent a step ahead in comparison to our previous proposals without the far-infrared analysis and without the most recent models of asphaltene molecules as reported by Mullins et al. (2007).

5- Infrared spectra of PAH mixtures and the 21 µm feature

Figure 5 shows that the BQ-1 asphaltenes show a relatively narrow band at about 21 µm without any other important feature in the far infrared up to 200 µm (see also Table 2). The same applies for the acene series of PAHs shown in Figure 6 and summarized in Table 3. In comparison to astronomical observations (e.g., Volk et al. 1999; Hrivnak et al. 2009), the 21 µm feature of BQ-1 asphaltenes and of acenes appear somewhat narrow and and slightly shifted to longer wavelengths. In order to broaden such feature, we have mixed together all the acenes in a 1:1:1:1 weight ratio. The infrared spectrum of the acene mixture is shown in Figure 10 (2nd spectrum from top to bottom) and is characterized by an infrared band at 21.24 µm with two sub-features at 20.75 and 21.88 µm. The addition of BQ-1 asphaltenes to the acene mixture (respecting the 1:1:1:1:1 weight ratio) causes a broadening of the infrared band which now appears centred at 21.20 μm. In addition, in order to increase to some extent the aliphatic and cycloaliphatic character of the mixture, the bitumen asphaltenes were also added. Figure 10 (3[rd] spectrum from top to bottom) shows a slight further band broadening with a peak centred at 21.19 μm. Simultaneously, it is possible to observe that the aliphatic/cycloaliphatic C-H stretching band at 3.5 μm is now more intense than the aromatic band at 3.3 μm, while in the absence of the bitumen asphaltenes the aromatic band at 3.3 μm was the most intense in the C-H stretching region (cfr. Fig. 10, 2[nd] from top with 3[rd] from top). This means that with our spectra we can change the ratio between the aromatic and aliphatic/cycloaliphatic bands in the C-H stretching region by changing the components of our mixture without affecting significantly the band pattern of the spectrum at longer wavelengths. This also maintains a broad feature at about 21 μm although, as we have noted above, the laboratory features do not match the characteristics (e.g., width and central wavelength) of the so-called 21 μm feature observed in PPNe (see Volk et al. 1999; Hrivnak et al. 2009).

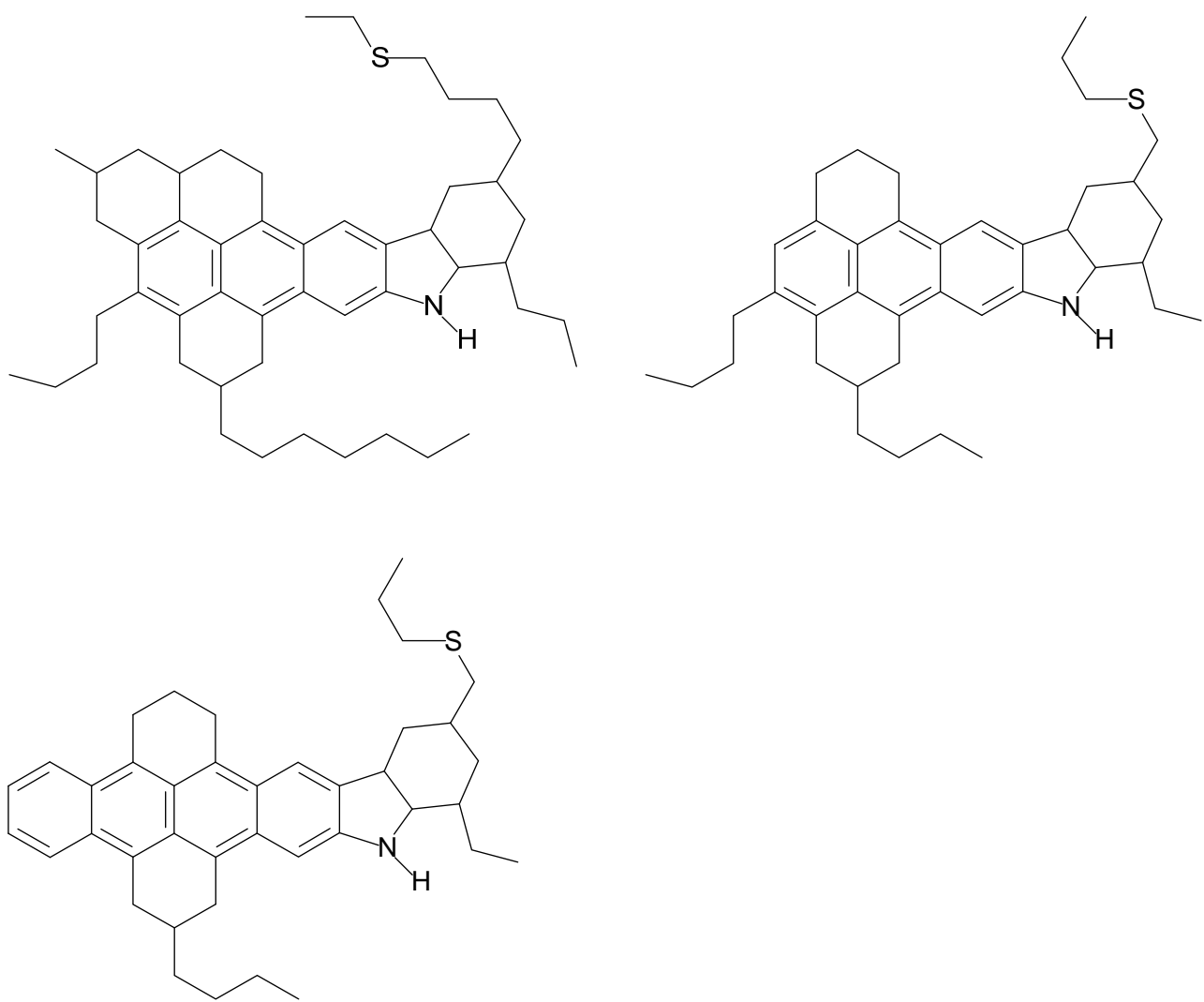

*Figure 9 – Petroleum asphaltene molecular models with an anthracene or tetracene "core" from BQ-1. The basic asphaltene structures were taken from Mullins et al. (2007) and modified slightly to take into account the current experimental data suggested by the far infrared spectra. These chemical structures represent the "average" possible structures of asphaltenes extracted from the petroleum fraction BQ-1.*

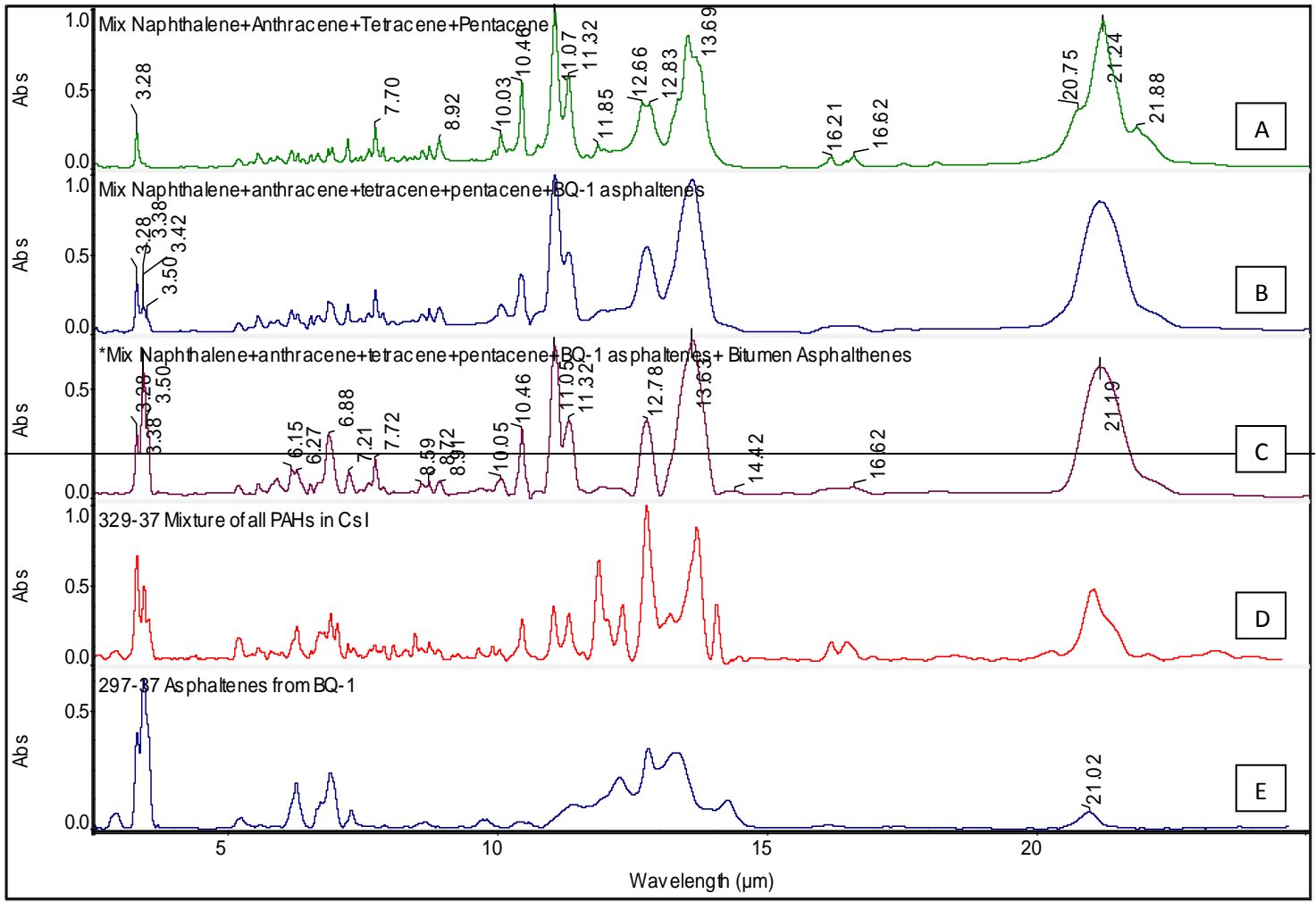

*Figure 10 – FT-IR spectra in the INFRARED between 2.6 and 25.0 μm from top to bottom: **A** = mixture of acenes (naphthalene+anthracene+tetracene+pentacene in the same weight proportions); **B** = mixture of acenes + BQ-1 asphaltenes fraction; **C** = mixture of acenes + BQ-1 asphaltenes and bitumen asphaltenes (in the same weight proportions); **D** = complex mixture of PAHs including acenes, phenanthrene and picene, pyrene, methylpyrene, perylene, acenaphtene, acenaphtylene and 9,10-dihydroanthracene (in the same weight proportions); **E** = BQ-1 asphaltenes fraction.*

The fourth infrared laboratory spectrum (from top to bottom) in Figure 10 is due to a complex mixture of PAHs, which includes not only the acenes but also other PAHs (see the legend below Figure 10). In these conditions, the infrared band at 21 μm appears less intense than in the other spectra and also asymmetrical in shape, confirming our original suggestion that we are dealing essentially with linearly condensed PAHs (i.e., acenes) without any important contribution from other types of PAHs having different anellation than acenes. Finally, the last spectrum at the bottom of Figure 10 is due to BQ-1 asphaltenes and its band pattern is matching only approximately that of the PAHs mixture shown above. However, the feature at about 21 μm is much less intense than that of the complex PAH mixtures mentioned before.

6 – Comparison with astronomical observations

A large variety of organic species have been observed to emerge in the circumstellar environment of stars in the transition phase between AGB stars and PNe (i.e., proto-PNe), following the formation of gas-phase molecules and inorganic solid-state compounds in the previous AGB phase. Contrary to much more complex astronomical environments (e.g., the diffuse interstellar medium, proto-planetary disks, reflection nebulae, galaxies, etc.), the main advantages of studying the circumstellar environment around proto-PNe (PPNe) lie in their well-determined physical conditions (e.g., density, temperature, radiation background), the singleness of the systems (a single energy source with a simple geometry) as well as the short chemical time scales that are defined by the dynamical time scales (only ~$10^3$ years for the proto-PNe phase). Thus, the well-defined conditions of these circumstellar environments approximate a controlled laboratory as closely as possible in space, providing severe constraints on any model of gas-phase and solid-state Chemistry and Physics (Kwok 2004). In addition, PPNe are the best objects for comparison with our materials because they usually display a mix of aliphatic and aromatic IR features (e.g., Kwok et al. 2001; Hrivnak et al. 2007; Kwok & Zhang 2011). It should be noted here, however, that we have also compared the characteristics (i.e., wavelength position) of the IR features seen in the laboratory spectra of the several materials presented here with other astronomical sources such as Reflection Nebulae and Galaxies and this comparison is summarized in Table 2.

It is worth to briefly mention that dust particles with mixed aromatic and aliphatic structures such as petroleum fractions, coal, and asphaltenes - if present in the circumstellar envelopes of PPNe - are supposed to be excited by stochastic heating (from UV to visible photons) similarly to PAHs and nanodust (see e.g., Li & Mann 2012 for a recent review). In addition, these aliphatic-aromatic materials could emit infrared emission as a consequence of the chemical energy model in HAC dust nano-particles recently proposed by Duley & Williams (2011).

Polycyclic aromatic hydrocarbons are formed in the outflow of carbon-rich stars by repeated addition of acetylenic units as summarized by Kwok (2012) and as earlier suggested by Frenklach & Feigelson (1989). Acetylenic units are the building blocks of polyynes chains which in their turn are cyclized into PAHs having unsaturated side chains under the form of acetylenic groups (Rouillé et al. 2012). If hydrogen is present, then the hydrogenation of the acetylenic side chains leads to PAHs with saturated alkyl chains substituents. The formation of PAHs and carbon nanoparticles with acetylenic side chains has been demonstrated also with laboratory experiments simulating the conditions of the circumstellar environment of late-type, carbon-rich stars (Duley & Hu 2009). Furthermore, the alkylation of PAHs has been demonstrated by Mahajan and colleagues (2003) to occur easily under a variety of UV irradiation conditions of mixtures of PAHs and hydrocarbons. Tielens (2005) has discussed about the stability toward UV irradiation of alkyl-substituted PAHs in comparison to normal PAHs once dispersed in the interstellar medium and came to the conclusion that long alkyl chain substituents are the most prone to be lost quite easily indicating the PAHs with very short side chains like methyl groups as those having a photochemical stability equivalent to that of normal PAHs. Till now the focus was on simple PAHs and little attention has been dedicated to alkylated PAHs which instead are very interesting molecular models after Kwok & Zhang (2012) have attracted our attention on the presence of organic matter with mixed aromatic/aliphatic and naphthenic character in the circumstellar medium of certain PPNe. This is the reason why we have selected both heavy petroleum fractions and used also the asphaltenes as molecular models to interpret the infrared emission spectra of certain PPNe. Indeed, heavy petroleum fractions exhibit a mixed aromatic/aliphatic and naphthenic chemical structure as already discussed, why asphaltenes are naturally occurring polyalkylated PAHs.

In order to compare the laboratory infrared spectra of the complex organic compounds presented in this paper (e.g., coal, heavy petroleum fractions, PAH mixtures, etc.) with the already available mid- and far-infrared observations of prototype astronomical sources, we have retrieved the ~2-200 μm infrared spectra of two PPNe (IRAS 22272+5435 and IRAS 07134+1005) and a young PNe (IRAS 21282+5050) from the Infrared Space Observatory (ISO) data archive. Thus, we normalize their infrared spectra by dividing the observed ISO spectra by the dust continuum emission, which was represented by five order polynomials fitted to spectral locations free from any dust or gas feature. These normalized PPNe/PNe spectra are compared to the laboratory data in Figures 11 to 13. It should be noted here that it is very unlikely that a detailed fit on the relative intensities between the laboratory spectra and the observational data could be obtained, but if the coincidence of the peak positions of the lab features with those in the PPNe data is significant, then this would indicate that laboratory samples may have a similar chemical composition to that of the dust in the circumstellar environment of PPNe and young PNe.

The weakness of the PAHs mixture model either in the acene version or in the complex PAHs mixture is shown in Figure 11, where the spectrum of the PAHs acene mixture (including also BQ-1 and bitumen) is compared with the PPNe IRAS 22272+5435 and IRAS 07134+1005 spectra. First, the band pattern between 5 and 15 µm of the PAHs mixture is so rich of sharp bands that is not matching even approximately the spectral pattern of the two PPNe. Furthermore, the intense far infrared band at 21.19 µm is very sharp and somewhat redshifted in comparison to the broad PPNe feature that is located closer to 20 µm. In Figure 11 is instead shown that the band pattern of DAE and RAE petroleum fractions but also of bitumen and coal are matching reasonably the infrared band pattern of the PPNe IRAS 22272+5435 and IRAS 07134+1005 spectra, confirming that in this spectral range the petroleum fractions and even coal models are more effective than the mixture of PAHs of any type. It is important to mention that the 20 µm feature of the PPNe spectra is not matched neither by the DAE and RAE fractions nor by the bitumen and coal. However, in Figure 12 it is evident that the petroleum fraction BQ-1 with its sharp band at 21.02 µm is not matching the position of the 20 µm feature in the PPNe spectra neither the band width. The merit of the BQ-1 petroleum fraction is that the band pattern of its spectrum in the 5-15 µm range is still matching in a reasonable way the spectra of PPNe and this is in contrast with the spectrum of the PAHs mixture discussed previously.

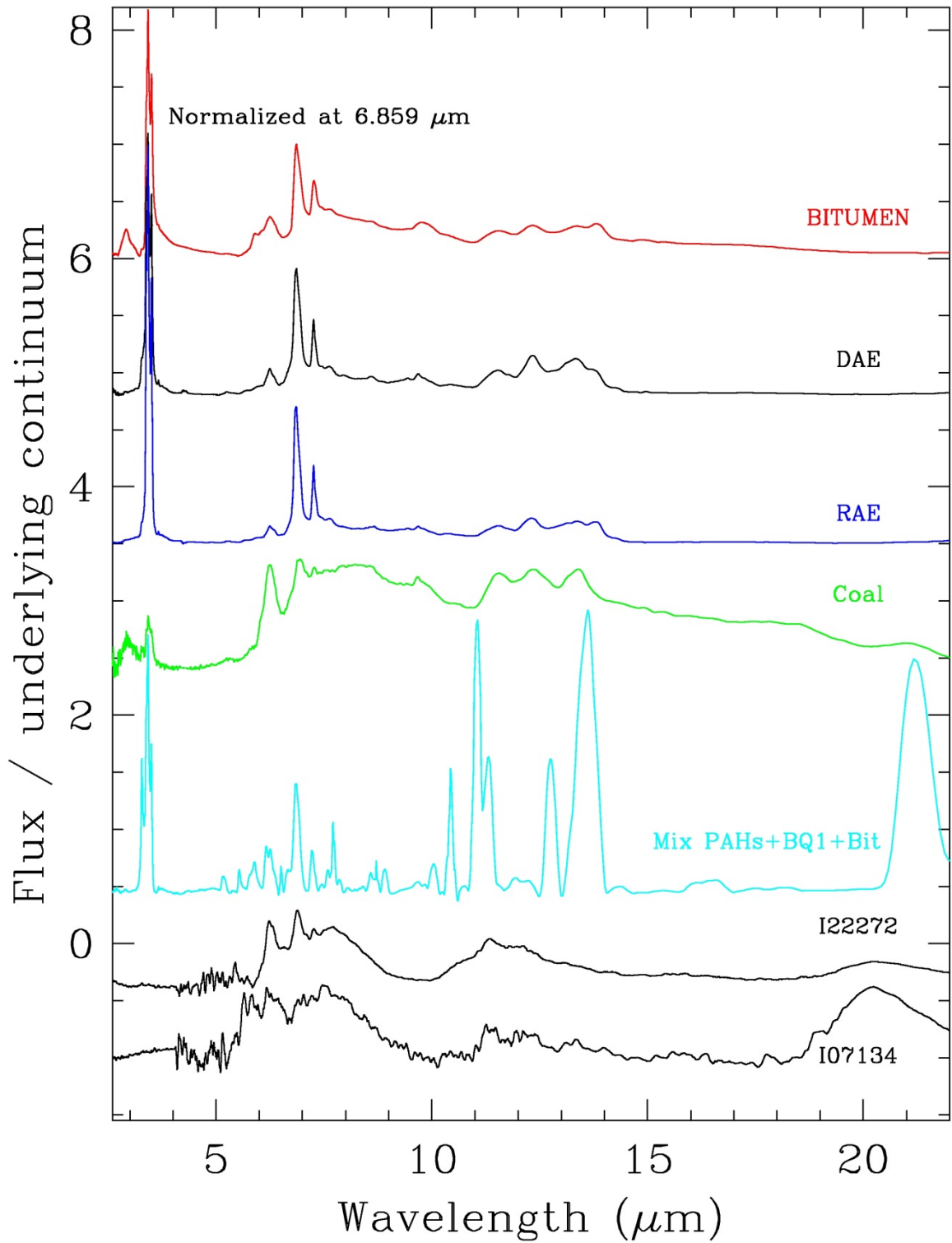

*Figure 11. Laboratory spectra for BITUMEN (in red), DAE (in black), RAE (in blue), Coal (in green) and mixture of acenes PAHs (in cyan) including also small amounts of BQ-1 and bitumen asphaltenes in the range ~3-22 μm in comparison with the Infrared Space Observatory spectra (Flux divided by the underlying continuum, see text) of the PPNe IRAS 22272+5435 and IRAS 07134+1005 (both in black). Note that the astronomical spectra have been smoothed (with a 75-point box car filter) to be compared with the laboratory spectra.*

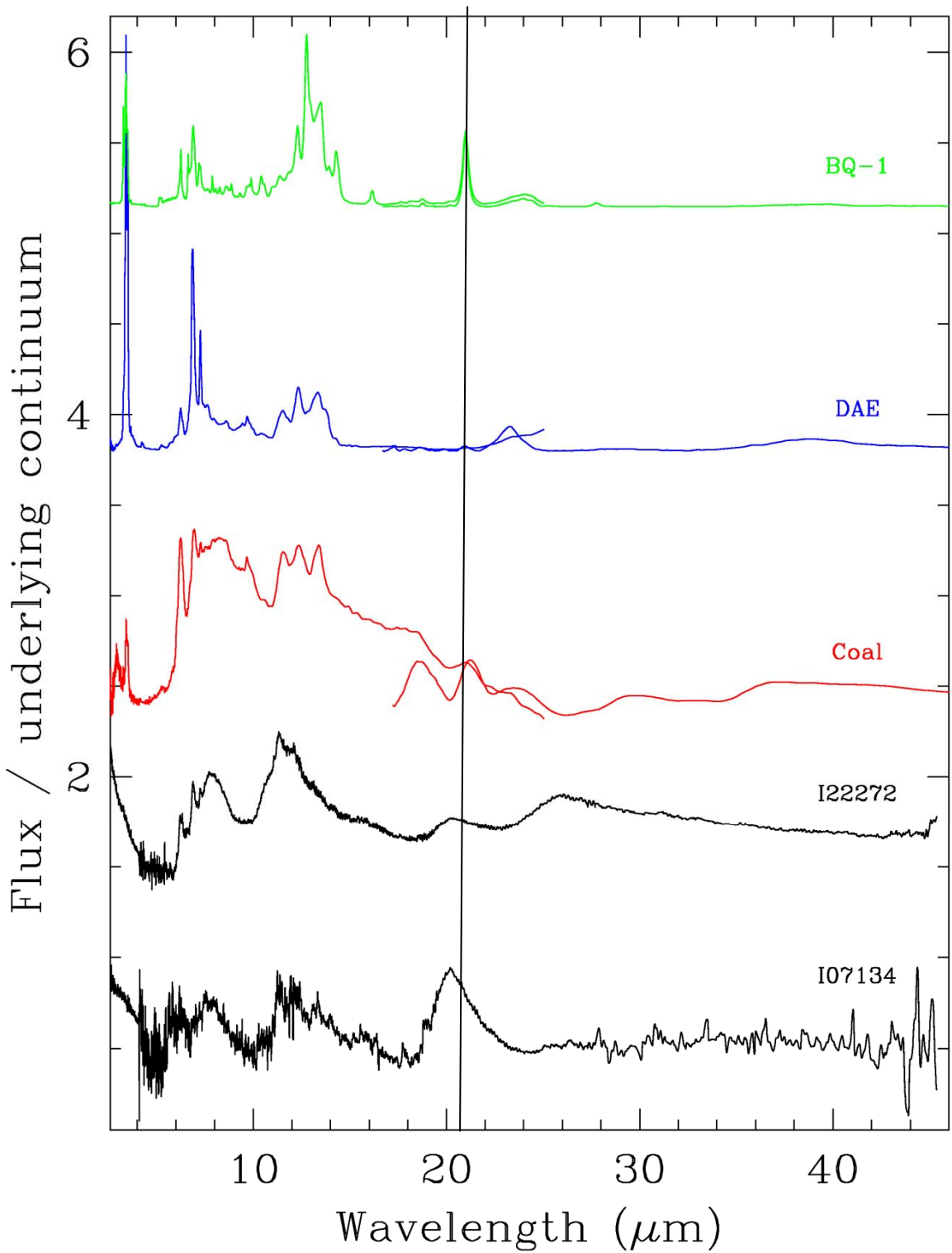

*Figure 12. Laboratory spectra for BQ-1 (in green), DAE (in blue), and coal (in red) in the range 2-50 μm in comparison with the Infrared Space Observatory spectra (Flux divided by the underlying continuum, see text) of the PPNe IRAS 22272+5435 and IRAS 07134+1005 (both in black). Note that this time the PPNe spectra have not been smoothed. Note that the far-infrared spectra were scaled to the mid-infrared ones (e.g., by using the 21 um feature that overlaps).*

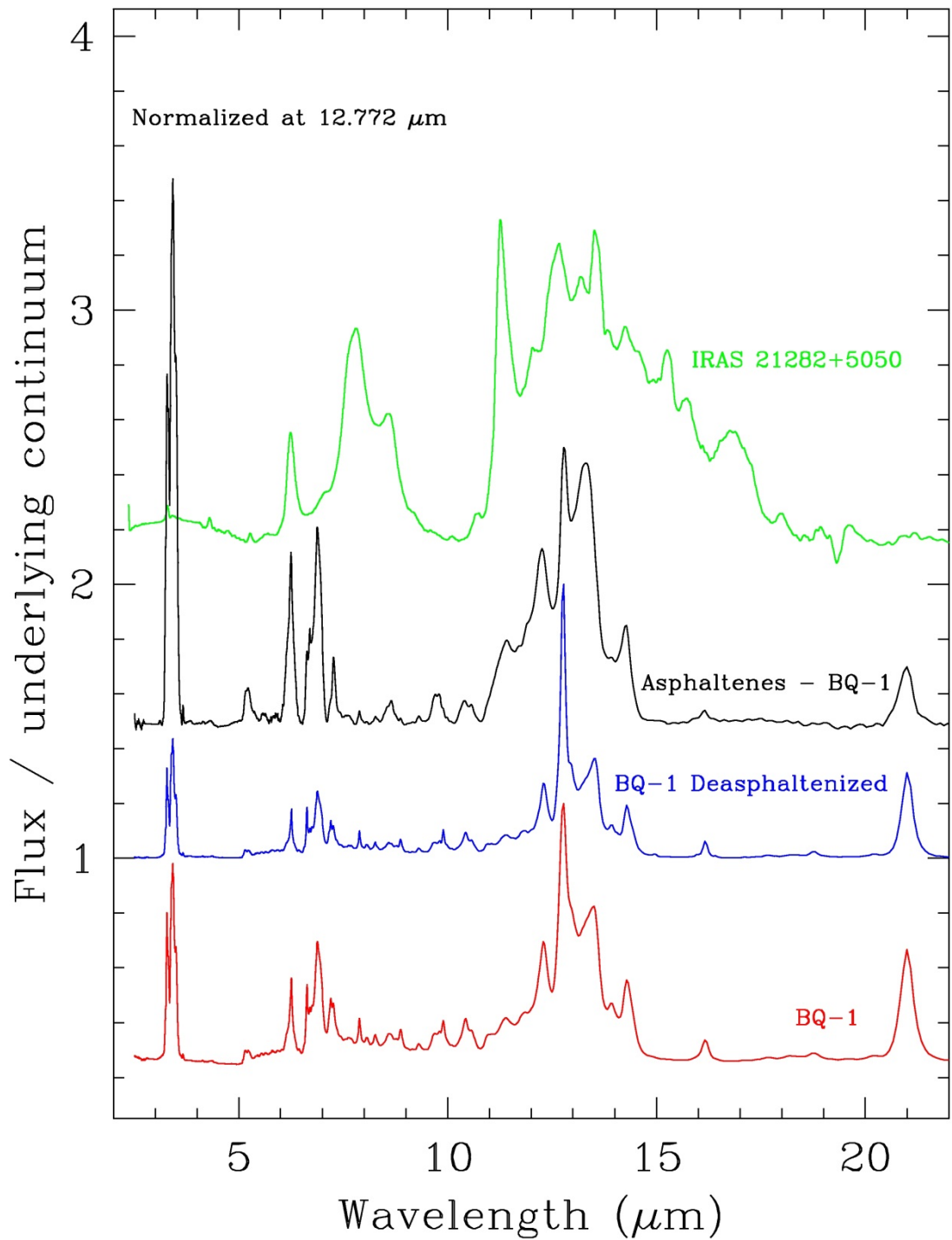

*Figure 13. Laboratory spectra for BQ-1 (in red), Deasphaltenized-BQ-1 (in blue), and Asphaltenes (in black) in the range ~3-22 μm in comparison with the Infrared Space Observatory spectra (Flux divided by the underlying continuum, see text) of the young PN IRAS 21282+5050 (in green). Note that the PN spectrum has been smoothed (with a 75-point box car filter) in order to be compared with to the laboratory spectra.*

Figure 13 is showing the incredibly good matching between the spectrum of the evolved young PN IRAS 21282+5050 and the BQ-1 petroleum fraction and its asphaltenes and deasphaltenized fraction. However, also some differences are evident: for example the IRAS 21282+5050 spectrum is showing also bands at 6.9 and 8.7 μm as well as at 16.95 μm, which do not have a counterpart in the BQ-1 spectra. Conversely, BQ-1 is showing the far infrared band at 21.02 μm without any similar feature in the IRAS 21282+5050 spectrum. In Table 2 are summarized all the mid- and far infrared bands of the selected petroleum fractions used in the present work, of coal and of the PPNe IRAS 22272+5435, the young PN IRAS 21282+5050, the reflection nebula NGC 7023 and from spectra taken in star forming galaxies regions. All the above data were taken from Kwok (2012).

## Conclusions

Among the various petroleum fractions, petroleum asphaltenes, coal, and bitumen used as reference materials the best match with the mid infrared spectra of the PPNe IRAS 22272+5435 and IRAS07134+1005 is offered by the petroleum fractions DAE and RAE and also by anthracite coal.

In the CH stretching region (~3-4 μm) the best match to the spectra of the PPNe IRAS 04296+3429, IRAS 22272+5435, and CRL 2688 is offered by the petroleum fraction BQ-1, which exhibits a mixed aromatic/aliphatic and cicloaliphatic character that is lacking in all the other petroleum fractions examined and also on coal. Only BQ-1 and its derivatives exhibit a prominent aromatic C-H stretching band in combination with the aliphatic C-H stretching band. BQ-1 and its asphaltenes are matching quite well the band pattern of the young PN IRAS 21282+5050.

A study of the petroleum fractions and coal in the far infrared has revealed that the petroleum fractions are made by a "core" of two or three condensed aromatic rings which are extensively alkylated by aliphatic and cycloaliphatic moieties. This conclusion is based on the fact that BQ-1 and derivatives display a simple far infrared spectrum characterized by a sharp peak at about 21 μm, analogous to the anthracene far infrared spectrum. Instead the other petroleum fractions, namely DAE and RAE exhibit a far infrared spectrum with a peak at 23.2 μm followed by a broad peak at about 39 μm, completely analogous to that of tetrahydronaphthalene. In contrast, the far infrared spectrum of coal is extremely rich of bands which are linked to its extremely complex crosslinked structure not yet fully well understood. This structural determination by far infrared spectroscopy is important because we can give and average chemical structure to our petroleum fractions models (see Figures 8 and 9) while the chemical structure of coal cannot be so easily modelized. Future observations of PPNe, PNe, and other astronomical objects in the far infrared (e.g., with the Herschel Space Observatory) will permit to establish if the petroleum fractions model is superior to that of the coal model. Indeed, far infrared spectra of astrophysical objects rich in a series of infrared bands will favor the coal model while a limited number of far infrared bands will favor the petroleum fractions model.

BQ-1 shows a sharp infrared band at about 21 μm and this should be compared with the 20 μm feature observed in different PPNe. Of course the BQ-1 feature is sharp while that of the PPNe is much broader. In order to broader the 21 μm feature, a mixture of acenes PAHs was prepared including or not including also BQ-1 and bitumen in the mixture. Indeed, a band broadening of the feature at 21.2 μm was obtained, however the infrared spectrum in the 5-15 μm range was not at all able to match the pattern of the PPNe spectra. This demonstrates that the PAHs mixture model is a very weak model while the BQ-1 petroleum fraction may contribute to the PPNe spectral feature at 20 μm but is cannot be used as model carrier of that spectral feature.

## Acknowledgements

The present research work has been supported by grant AYA2007-64748 Expte. NG-014-10 of the Spanish Ministry of Science and Innovation (MICINN). D.A.G.H. and A.M. acknowledge support provided by the Spanish Ministry of Economy and Competitiveness under grant AYA2011-27754. The authors wish to thank Prof. Sun

Kwok for the stimulating discussions about the far infrared spectra of asphaltenes and petroleum fractions and their potential significance in the interpretation of the emission spectra of certain PPNe.